\documentclass[doublespacing]{elsart}

\usepackage{graphics}
\usepackage{epsfig}

\usepackage{amssymb}

\begin{document}

\begin{frontmatter}



\title{Accurate measurement of the electron beam polarization in JLab Hall A using Compton polarimetry
       \thanksref{aknow}}
\thanks[aknow]{We wish to thank the JLAB-hall A technicians and the Accelerator Operation group for their critical help and support in the operation of the Compton polarimeter. This work was supported by the Commissariat \`a l'\'Energie Atomique and by DOE contract DE-AC05-84ER40150, modification No. M175, under which the Southeastern Universities Research Association (SURA) operates the Thomas Jefferson National Accelerator Facility.}


\author[cea]{S. Escoffier\thanksref{presad1}}, \author[lpc]{P.Y.~Bertin}, \author[lpc]{M.~Brossard}, \author[cea]{E.~Burtin}, \author[cea]{C.~Cavata}, \author[cea]{N.~Colombel}, \author[jlab]{C.W.~de~Jager},
\author[cea]{A.~Delbart}, \author[cea]{D.~Lhuillier\corauthref{cora}}, \author[cea]{F.~Marie},
\author[jlab]{J.~Mitchell\thanksref{presad2}}, \author[cea]{D.~Neyret} and \author[cea]{T.~Pussieux}

\corauth[cora]{Corresponding author. ph: 33-1-69-08-94-97, fax: 33-1-69-08-75-84, email: dlhuillier@cea.fr}
\thanks[presad1]{CNRS/IN2P3/Centre de Physique des Particules de Marseille, 163 avenue de Luminy, 13288 Marseille cedex 09, FRANCE}
\thanks[presad2]{Present address: Renaissance Technologies Corp., Setauket, NY 11733, USA}

\address[cea]{
CEA Saclay
DSM/DAPNIA
F-91191 Gif-sur-Yvette Cedex
France}

\address[lpc]{
Universit\'e Blaise Pascal et CNRS/IN2P3 LPC
6F-3177 Aubi\`ere Cedex
France}

\address[jlab]{
Jefferson Lab
12000 Jefferson Avenue
Newport News, VA 23606
USA}

\begin{abstract}
A major advance in accurate electron beam polarization measurement has been achieved at Jlab Hall A with a Compton polarimeter based on a Fabry-Perot cavity photon beam amplifier. At an electron energy of 4.6~GeV and a beam current of 40~$\mu$A, a total relative uncertainty of 1.5\% is typically achieved within 40~min of data taking. Under the same conditions monitoring of the polarization is accurate at a level of 1\%. These unprecedented results make Compton polarimetry an essential tool for modern parity-violation experiments, which require very accurate electron beam polarization measurements.
\end{abstract}

\begin{keyword}
Compton polarimeter \sep optical cavity \sep polarized beams
\PACS 07.60.Fs \sep 42.60.-v \sep 29.27.Hj

\end{keyword}
\end{frontmatter}

\section{Introduction}
The Continuous Electron Beam Accelerator Facility (CEBAF) at the Jefferson Laboratory(JLab) is a new particle accelerator which makes extensive use of its highly polarized electron beam for the study of nucleons and nuclei. The polarization is measured at the injector with a 5~MeV Mott polarimeter and in the Hall A beam line with a M{\o}ller polarimeter and a Compton polarimeter. Mott and M{\o}ller polarimeters require solid targets and operate respectively at low energy and at low intensity ($\approx 1\mu$A). Because of its thin "photon target", only Compton backscattering polarimetry provides an essential tool for accurate measurement and monitoring of the beam polarization under the same conditions as the running experiment. However, the mean Compton analyzing power ($A_c$) depends strongly on the beam energy - $A_c\approx 0.4$\%/GeV - while the total cross section is approximately constant at 0.6 barn. Therefore, the typical beam conditions provided by CEBAF, an energy of several GeV and a beam intensity up to 100 $\mu$A, require a high laser power to provide the required interaction luminosity. The design of the Compton polarimeter was challenging \cite{CDR,Jorda}. The photon density is amplified with a Fabry-Perot cavity of very high finesse which provides a power of 1700~W of IR light at the Compton interaction point. This performance, unequalled in a particle accelerator environment, results in a statistical accuracy for a polarization measurement below 1\% within an hour at 4.6~GeV~\cite{Nico}.  This number scales with the inverse of the beam energy.\\
In section~\ref{sec:cptjlab} of this paper, we briefly summarize the experimental set-up of the Compton polarimeter. Section~ref{sec:datatake} describes its operational properties achieved during two polarized experiments, $N-\Delta$~\cite{NDelta,Roche} and GEp~\cite{GEp,Gayou}. Next, we describe a new analysis method developed to restrain systematic uncertainties in the polarization measurement with a high confidence level. We explain in detail the sources of these systematic errors and present longitudinal electron polarization measurement results. Finally, we give for the first time at JLab a measurement of the polarization difference between the two helicity states of the electron beam.

\section{Compton polarimeter at JLab}
\label{sec:cptjlab}
Compton scattering of polarized electrons off a circularly polarized photon beam shows an asymmetry of the counting rates $n^{+/-}$ for different orientations of the electron polarization~\cite{Prescott}
\begin{eqnarray}
A_{exp}&=&\frac{n^+-n^-}{n^++n^-}\,=\,P_e P_\gamma A_c
\end{eqnarray}
where the asymmetry $A_c$ is calculated from QED. Measurements of the experimental asymmetry $A_{exp}$ and of the circular photon polarization $P_\gamma$ give access to the mean longitudinal electron polarization $P_e$. The electron beam polarization is flipped at a 30~Hz rate to minimize systematic effects.\\
The Compton polarimeter is composed of a magnetic chicane of four identical dipoles connected in series and installed in the Hall A beam line. The Compton interaction takes place at the center of a symmetric Fabry-Perot cavity in which photons, originating from a 230~mW IR laser ($\lambda= 1064$~nm) interfere. The laser frequency is locked to one of the resonant frequencies of the cavity using the Pound-Drever feedback technique~\cite{Drever}. The maximum power inside the cavity reaches 1700~W, with a coupling to the fundamental mode of 92\%. The detectors are installed between the third and the fourth dipoles of the chicane. The backscattered photons go through the gap of the third magnet in a  calorimeter consisting of 25 $PbWO_4$ crystals (2x2x23cm$^3$) and the scattered electrons are detected in 4 planes of 48 silicon strips (650~$\mu$m wide), segmented along the dispersive axis. The data acquisition can be triggered by either electrons, photons, or both (in coincidence). Typical running conditions at CEBAF during data taking in 2000 were an electron energy of E=4.6~GeV and a beam current of $I_e=40\mu$A. The Compton backscattered photons' energy range is from 0 to 340~MeV under these conditions.

\section{Data Taking}
\label{sec:datatake}
We describe here how the Compton polarimeter data-acquisition system works, and the strategy used to minimize false asymmetries.
\subsection{Acquisition}
The data acquisition is driven by the 30~Hz electron beam polarization flip. Two milliseconds after each reversal, the trigger system is activated and events are taken from the photon and/or electron detectors, according to the trigger configuration determined by the user. The trigger system is inhibited a few ms before the next reversal.\\
Each detector has its own trigger logic. The photon calorimeter trigger system generates an event when the signal of one the photo-multiplier tubes exceeds a given threshold. This signal is then integrated over a period of 150~ns. The electron detector triggers when signals are detected in coincidence on a given number of the silicon strip planes, at the same dispersive position. A specific logic is used to take care of cases where both detectors fire in coincidence.\\
The data-acquisition system can read out photon and electron events at a rate greater than 100~kHz with a dead time of only a few percent. These data are read by either a custom-built buffer card for the electron detector signals, or 10 bits buffered ADCs for the photon calorimeter. Calibration signals from a LED can be used to monitor the gain variation of the photon detector.\\
All these raw data are read through VME block transfer by two Power PC CPU cards working alternatively at each electron polarization reversal. At the end of each polarization period, the CPU card that has read out the data, reads values from scaler cards which provide summary information of that period (counting rates, number of triggers, dead time, average value of electron and photon beams parameters, etc\dots). This CPU then transfers control of the VME crate to the other CPU, produces on-line calculations and sends a data block to a workstation where these data are stored. The goal of these on-line calculations is to reduce the huge amount of data coming from both detectors by producing computed values and histograms (in particular the energy spectra of the scattered photons). Only a small fraction of the raw data, controlled by prescaler factors, is kept for monitoring purposes. Thus, the data block stored at the end of each electron polarization state consists of the scalers' summary values, the result of the on-line calculations (computed values and histogram), and pre-scaled photon, electron and coincidence raw data.

\subsection{Photon polarization reversal}
\begin{figure}[t]
\begin{center}
  \scalebox{0.6}{
    \includegraphics{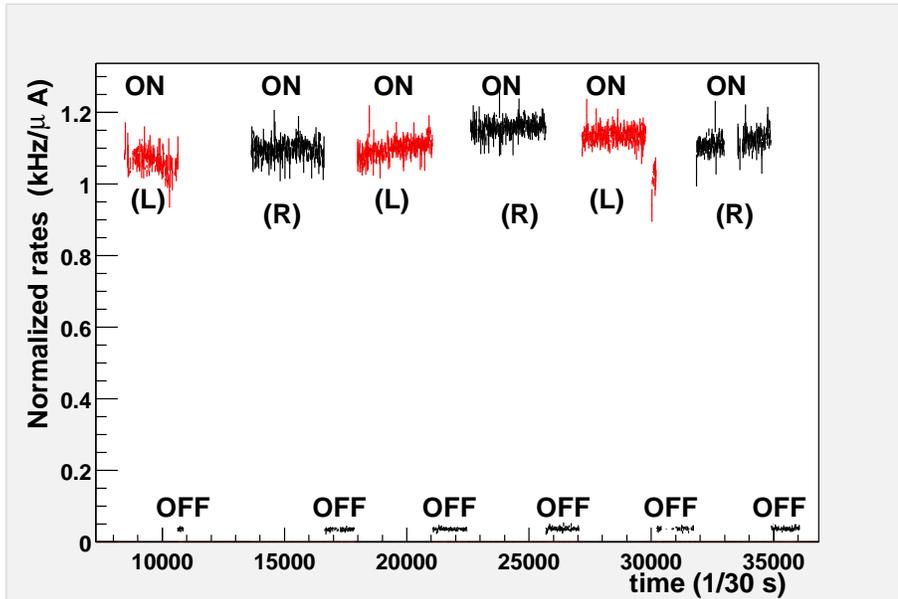}
  }
\end{center}
\caption{\label{fig:OnOff} Normalized counting rates versus time with alternate Left (L) and Right (R) circular polarization of the photon separated by laser OFF periods to monitor the background level.}
\end{figure}

Helicity-correlated differences in the electron beam parameters (charge, position and angle) lead to false asymmetries $b_i$ which add to the experimental asymmetry
\begin{equation}
\label{eq:Aexp}
A_{exp}\,=\,P_e P_\gamma A_c + \sum_i b_i
\end{equation}
where $i$ runs over the different sources of false asymmetries.
The charge asymmetry is corrected to first order by normalizing the counting rates to the beam current. The remaining systematic effects from position and angle are independent of the photon beam polarization state. Hence, in changing the sign of the photon polarization the major part of this type of false asymmetries is canceled.
This defines the procedure for data taking as a sequence of alternating right and left laser circular polarization, as illustrated in figure~\ref{fig:OnOff}. Moreover, between two photon polarization states, the cavity is unlocked in order to measure the background. Thanks to a high quality vacuum in the beam pipe and the control of the beam envelope using quadrupoles upstream the magnetic chicane a signal over background ratio of 20 is routinely achieved.

\section{Experimental asymmetry}
\begin{figure}[b]
\begin{center}
  \scalebox{0.6}{
    \includegraphics{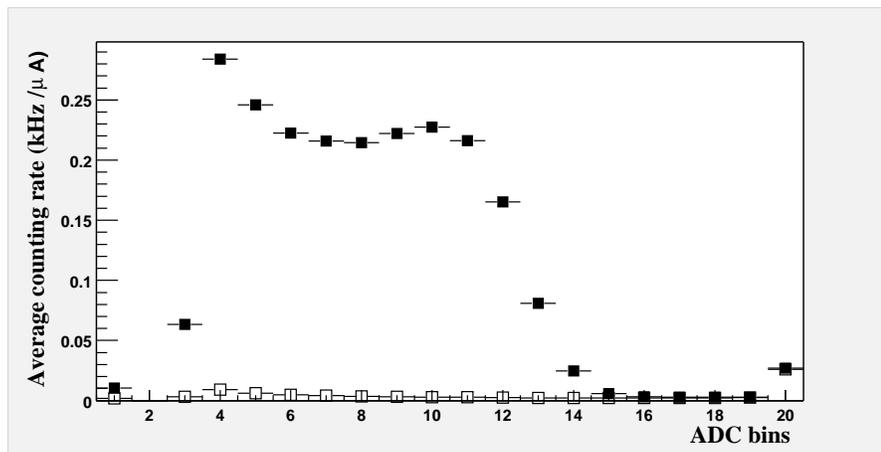}
  }
\end{center}
\caption{\label{fig:NormRate} Normalized counting rates versus ADC bins of the energy deposited in the photon  calorimeter, for laser ON (solid squares) and laser OFF (empty squares) periods.}
\end{figure}

\begin{figure}[t]
\begin{center}
  \scalebox{0.5}{
    \includegraphics{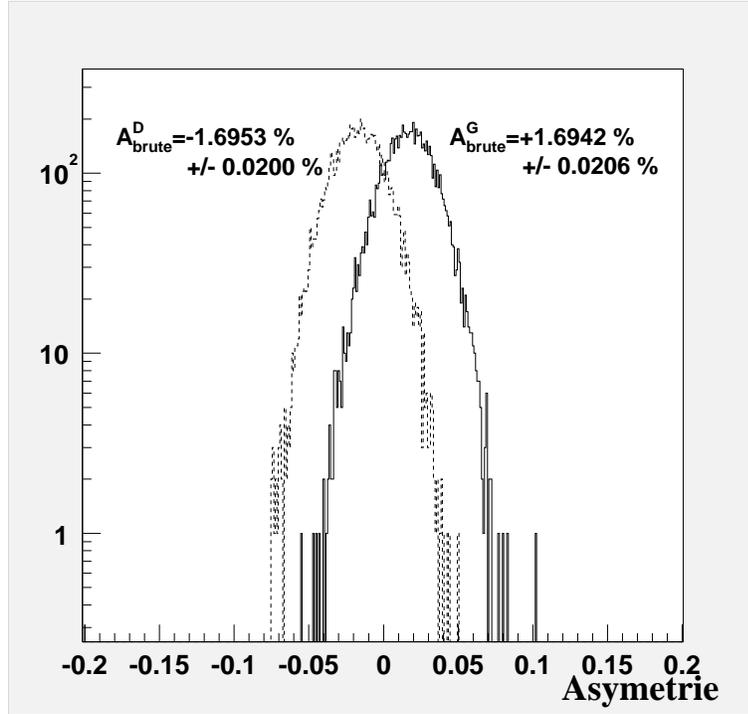}
  }
\end{center}
\caption{\label{fig:ApDist} Distribution of pulse-to-pulse asymmetries $A_p$ for both right and left photon polarizations. The same size but opposite sign of the mean values is a check of systematic effects.}
\end{figure}

For a given circular photon polarization, right (R) or left (L), we can calculate the asymmetry of integrated event numbers for two consecutive windows of opposite electron helicity states, as
\begin{equation}
A_p^{R/L}\,=\,\frac{n^+_{R/L}-n^-_{R/L}}{n^+_{R/L}+n^-_{R/L}}
\end{equation}
where $n^\pm$ refers to the normalized numbers of photons with a deposited energy greater than a given threshold.  These are defined as
\begin{equation}
n^\pm\,=\,\frac{\sum_{i>i_s}N_i^\pm}{I^\pm \Gamma^\pm}
\end{equation}
where $I\pm$ is the electron beam intensity, $\Gamma^\pm$ is the acquisition live time, $N_i^\pm$ is the number of detected events in the $i^{th}$ ADC bin and $i_s$ is the threshold corresponding to the lower edge of the bin. The normalized counting rates $N\pm/I^\pm\Gamma^\pm$ are shown in figure~\ref{fig:NormRate} versus the
energy in ADC bin units. The threshold $i_s$ is a software threshold applied to the total charge deposited and not to the maximum amplitude
reached by the signal. It can be varied off line in order to obtain the optimal value that maximizes the statistical accuracy and minimizes
the effect of false asymmetries. This operating point is found to be between the 6th and the 9th bin (see section~\ref{sec:syst}).
For a typical 40 minutes run, a raw asymmetry $A_{raw}^{R/L}$ is defined as the average of all pulse-to-pulse asymmetries $A_p^{R/L}$. The distribution of these asymmetries is shown in figure~\ref{fig:ApDist}, for both right and left photon polarizations. We can see that the pulse-to-pulse asymmetry distributions follow a Gaussian law.
The raw asymmetry has to be corrected for background according to
\begin{equation}
A_{exp}^{R/L}\,=\,\left[ 1+\frac{B}{S}^{R/L}\right] A_{raw}^{R/L} - \frac{B}{S}^{R/L} A_B
\end{equation}
where $(B/S)^{R/L}$ is the background to signal ratio for each photon polarization and $A_B$ is the background asymmetry. $B/S$ is of the order of 0.06 with a threshold set to the $8^{th}$ energy bin ($\approx 230$~MeV), and $A_B$ is found to be compatible with zero at the $10^{-4}$ level.\\
Finally, the mean experimental asymmetry is computed as
\begin{equation}
\label{eq:AexpOm}
\left<A_{exp}\right>\,=\,\frac{\omega^L A_{Exp}^L-\omega^R A_{Exp}^R}{\omega^L+\omega^R}
\end{equation}
where $\omega^{R/L}$ corresponds to the statistical weight of each experimental asymmetry.\\
The mean experimental asymmetries measured above the software threshold for $E=4.6$~GeV are around 6\% and can be measured with a relative statistical accuracy of 0.65\% in one hour at $I=40\mu$A.

\section{Analysing power}
The second part of this analysis concerns the determination of the analyzing power. In order to account for detection effects, we define the response function of the calorimeter $\mathcal{R}(ADC, k)$ as the ADC spectrum for a set of photons with a given energy $k$. From this response function the probability to detect photons of energy $k$ above a given ADC threshold $ADC_s$ is
\begin{equation}
\label{eq:Pofk}
P(k)\,=\,\frac{\int^\infty_{ADC_s}\mathcal{R}(ADC, k)\,dADC}{\int^\infty_0\mathcal{R}(ADC, k)\,dADC}
\end{equation}
Using this probability one can then calculates the analyzing power of the polarimeter defined as the average of the Compton asymmetry weighted by the Compton cross section
\begin{equation}
\label{eq:AnaP}
\left<A_s\right>\,=\,\frac{\int^{k_{max}}_0 P(k)\,\frac{d\sigma_0}{dk}\,A_C(k)\,dk}
                        {\int^{k_{max}}_0 P(k)\,\frac{d\sigma_0}{dk}\,dk}
\end{equation}
\subsection{Determination of the response function $\mathcal{R}(ADC, k)$}
\label{sec:response}
\begin{figure}[b]
\begin{center}
  \scalebox{0.6}{
    \includegraphics{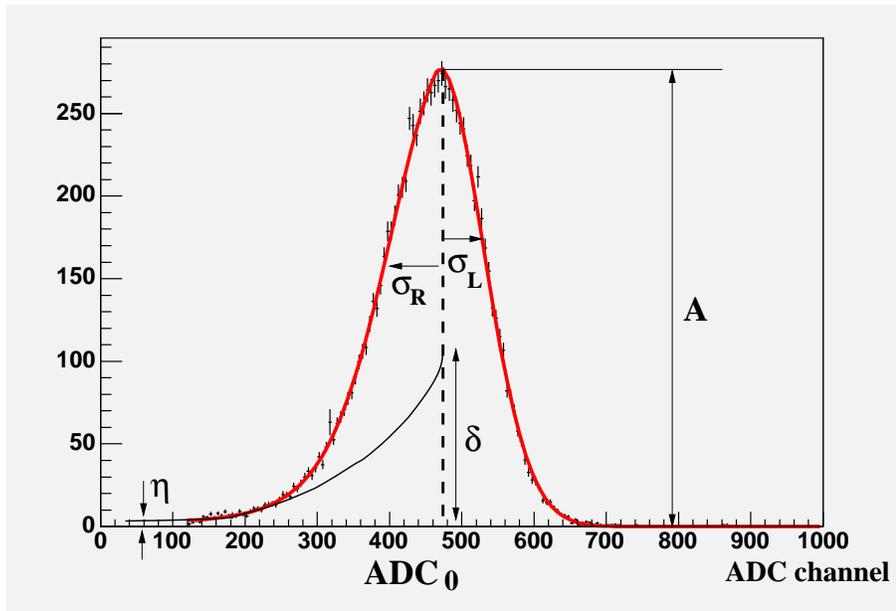}
  }
\end{center}
\caption{\label{fig:Espectrum} Photon energy spectrum measured in coincidence with electrons hitting the $14^{th}$ strip. Parameters of the fitting function are illustrated.}
\end{figure}

The calorimeter response function depends mostly on the intrinsic properties of the calorimeter. It is measured during dedicated runs where data are taken in photon-electron coincidence mode on an event-by-event basis.\\
Thanks to its very fine pitch the electron detector functions as an energy tagger of the incident photons. The distribution of the photon energy deposited in the central crystal for one selected strip of the electron detector is shown in figure~\ref{fig:Espectrum}. The tail at low energy is due to shower leakage to the sides of the central crystal (the Moli\`ere radius is 2.19 cm). For practical reasons it was found more accurate to model the response function of the central crystal rather than dealing with the inter-calibration of all the crystals of the 5x5 matrix~\cite{Maud}.
The response function is described by an {\it ad hoc} asymmetrical function composed of two Gaussians and a $4^{th}$ degree polynomial $P_4(x)$. Best fits were obtained with the following fit function
\begin{eqnarray}
\mathcal{R}(ADC, k)&=& A\,e^{\frac{(ADC-ADC_0)^2}{2\sigma_R^2}},\qquad ADC\geq ADC_0\\
\mathcal{R}(ADC, k)&=& A\,\left[ (1-\delta)\,e^{\frac{(ADC-ADC_0)^2}{2\sigma_L^2}}
                                +\eta +(\delta-\eta)\frac{ADC^4}{ADC_0^4}\right],\,ADC\leq ADC_0\nonumber
\end{eqnarray}
where $A$, $ADC_0$ and $\sigma_{R/L}$ are Gaussian parameters, and $\eta$, $\delta$ denote proportional amplitudes $P_4(0)/A$ and $P_4(x_0)/A$, as described in figure~\ref{fig:Espectrum}. $A$ is fixed by normalizing the integral of the response function to 1 in the denominator of Eq.(\ref{eq:Pofk}).
The remaining five parameters are functions of the scattered photon energy $k$, fitted to data from all electron detector strips which fired.
The Gaussian widths $\sigma_{R/L}$ are corrected for smearing due to the width of the electron strips ($\sigma_E\approx 5$~MeV).\\
The electron detector cannot be put closer than a few mm to the beam axis and thus restricts the range over which the response function can be determined. For instance, only photon energies between 150~MeV and 340~MeV (Compton edge) could be explored with a 4.6~GeV beam. The determination of the calorimeter response function is well controlled inside this energy range but the extrapolation to lower energy induces larger systematic errors (see section~\ref{sec:syst}).\\
\subsection{Calibration and analyzing power}
\begin{figure}[t]
\begin{center}
  \scalebox{0.6}{
    \includegraphics{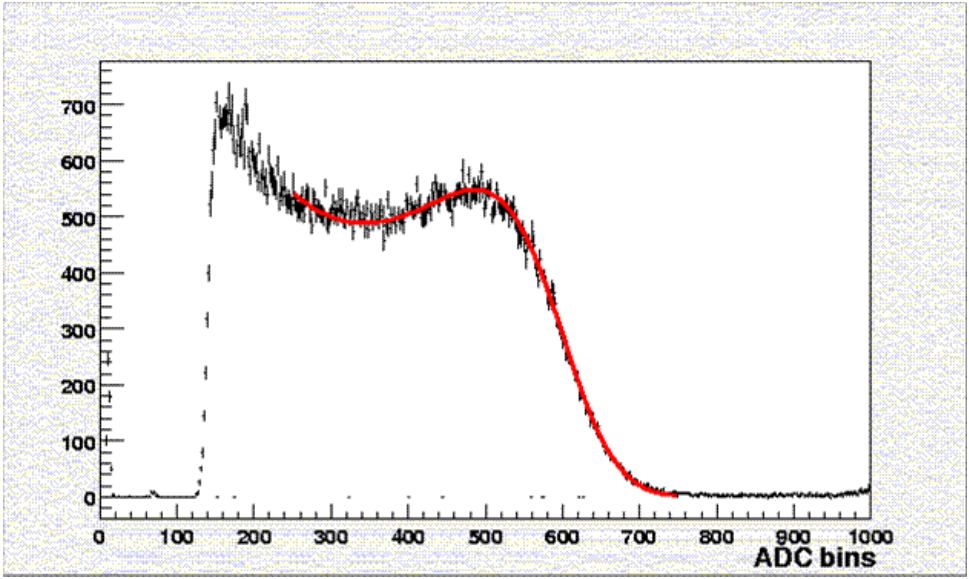}
  }
\end{center}
\caption{\label{fig:ExpFit} Fit of the experimental photon spectrum using the smeared cross section. The fit range is restricted to the validity energy range of the modelling.}
\end{figure}

\begin{figure}[b]
\begin{center}
  \scalebox{0.6}{
    \includegraphics{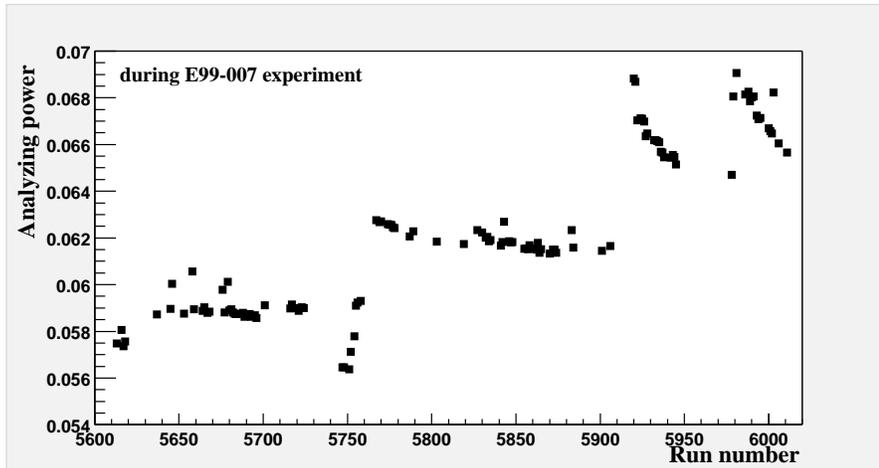}
  }
\end{center}
\caption{\label{fig:AnaP} Analysing power for each Compton run during the $G_E^P$ experiment.}
\end{figure}

The response function measured during a specific reference run has to be corrected for mean gain
variations when used to analyze a later run. To this end a calibration coefficient $\lambda$ is
introduced which accounts for gain corrections
\begin{equation}
\mathcal{R}(ADC, k)\,=\,\mathcal{R}(\frac{ADC}{\lambda}, k)
\end{equation}
$\lambda$ is fitted to the experimental spectrum of each run (Fig.~\ref{fig:ExpFit}) using the
convolution of the unpolarized Compton cross section $d\sigma_0(k)/dk$ with the response function
\begin{equation}
\frac{dN(ADC)}{dADC}\,=\,\int_0^{k_{max}}\frac{d\sigma_0(k)}{dk}\,\mathcal{R}(ADC, k)\,dk
\end{equation}
The probability of photon detection is deduced from Eq.(\ref{eq:Pofk}), where the lower integration
boundary $ADC_s$ is replaced by $ADC_s/\lambda$.
The analyzing power is then calculated from Eq.(\ref{eq:AnaP}) for each data run (with $i_s=8$).
An overview is given in figure~\ref{fig:AnaP} and shows relative variations of up to 10\%. Most of the
observed steps in the analyzing power originate from a retuning of the photon detection gain (PMT high
voltage or gain of the amplifier). "Reference" runs are repeated every few hours to check the consistency
of the extracted response function.

\section{Systematic uncertainties}
\label{sec:syst}
\subsection{Experimental asymmetry}
The largest source of systematic error in the experimental asymmetry is the false asymmetry related to the electron beam position, since the Compton luminosity is determined by the overlap of the electron and laser beams. If one assumes a Gaussian intensity profile for these two beams, the luminosity is also a Gaussian function of the distance between the two beam centroids. Since the optical axis of the cavity is fixed by the monolithic mechanic of the mirror holder, the position variation of the electron beam directly affects the Compton luminosity with a sensitivity equal to the derivative of this Gaussian function. In order to minimize this effect, two position-feedback systems were used, one at high frequency to reduce the jitter (down to 20~$\mu$m) and one at low frequency to lock the mean position at the point corresponding to the maximum of the Gaussian overlap curve, where the sensitivity to beam position goes to zero.
Finally, averaging over several photon polarization reversals cancels out most of these false asymmetries provided that the statistical weights of right and left circularly photon polarization states are similar. In practice, these statistical weights $\omega^{R/L}$ are not exactly equal, and some residual effects must be taken into account. So, in agreement with equations (\ref{eq:Aexp}) and (\ref{eq:AexpOm}), we have:
\begin{equation}
\label{eq:ASysExp}
\Delta A^{syst}_{exp_{i}}=res(b_i)=\frac{\omega^{L} b_i^L-\omega^{R} b_i^R}{\omega^{L}+\omega^{R}}
\end{equation}
Studies of the four beam parameters (x, y, $\theta_x$, $\theta_y$) show that their correlations tend to reduce the total false asymmetry. As a safe and simpler estimate of the error we assume them to be uncorrelated. The final error quoted in Table~\ref{tab:ExpSyst} should be read as a typical run-to-run error.  It corresponds to the width of the distribution of all $res(b_i)$ which turns out to be centered at zero. For each individual run one can also choose to correct for $res(b_i)$ and its error. When averaging the
 polarization over a sufficient number of runs $N_r$ the two approaches are equivalent and the systematic error reduces as $1/\sqrt{N_r}$.
The measured background has a small rate and asymmetry, compared to the Compton process, resulting in a negligible systematic error. Similarly the beam current asymmetry is at the few 100~ppm level and does not affect significantly the Compton asymmetry. The correction for the acquisition dead time is checked by comparing the scaler asymmetry and the corrected ADC asymmetry without applying a software threshold.
\subsection{Analyzing power}
\begin{figure}[t]
\begin{center}
  \scalebox{0.6}{
    \includegraphics{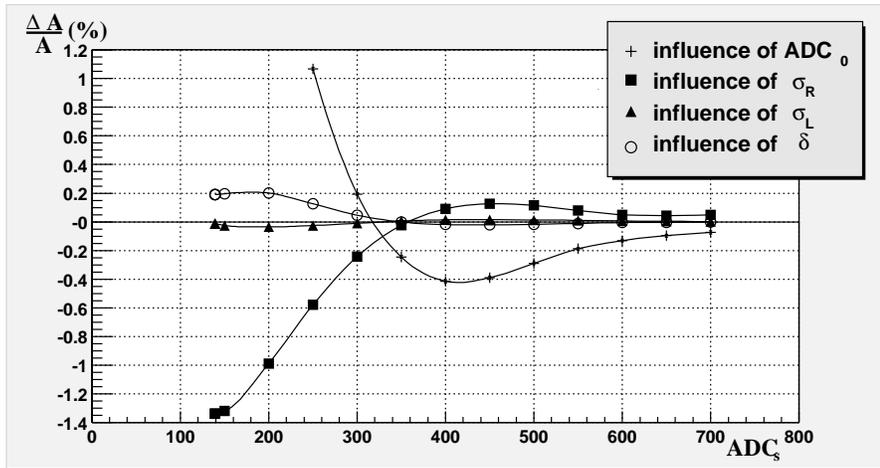}
  }
\end{center}
\caption{\label{fig:SystModel} Relative effects on analyzing power due to modeling of response function parameters, versus ADC threshold.}
\end{figure}

\begin{figure}[b]
\begin{center}
  \scalebox{0.6}{
    \includegraphics{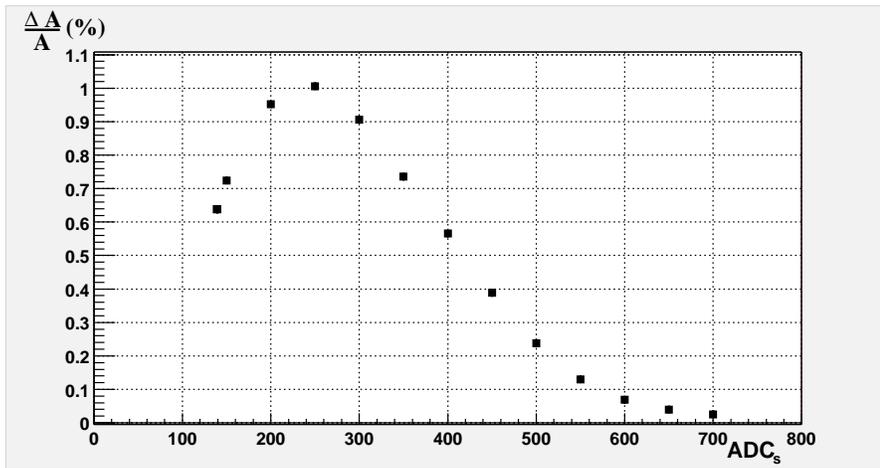}
  }
\end{center}
\caption{\label{fig:SystCalib} Relative deviation of the analyzing power due to the calibration error of the electron detector.}
\end{figure}

There are four main sources of uncertainties in the analyzing power. The first comes from the dependence of the response function on the parameterization used to describe it. To compute the systematic error we look at the variation of the analyzing power for a set of parameterizations with equivalent $\chi^2$ and try to define an envelope (fig. \ref{fig:SystModel}). For a threshold taken around ADC = 400 ($i_s = 8$, $E =230$~MeV) the effect is less than 0.45\%. Note that below the electron cut, located around channel 260 on fig.~\ref{fig:SystModel}, the systematic errors increase steeply. The second source of uncertainty arises from the momentum calibration of the electron detector which is used as an energy tagger in the determination of the response function (section~\ref{sec:response}). This calibration error is due to the position resolution of the electron detector (200~$\mu$m). The impact of this effect on the analyzing power is shown in figure~\ref{fig:SystCalib} as a function of the ADC threshold. For a threshold taken at ADC = 400 the effect is 0.6 \%. The third uncertainty is due to pile-up when two events are detected within the same acquisition gate ( $\sim$~150 ns) and are recorded as a single event of higher energy. The Compton spectrum is then shifted to higher energies. This modifies not only the experimental asymmetry but also the analyzing power via the calibration coefficient $\lambda$. Monte-Carlo simulations~\cite{Cpt_Steffie}
were performed for a measured pile-up rate of 0.9\%. They show a relative effect of 0.45\% for an ADC threshold $i_s = 8$. The fourth systematic uncertainty is due to the radiative corrections in real Compton scattering. The interfering process $e^- \gamma \rightarrow e^- \gamma \gamma$ causes a deviation of the analyzing power by about 0.26\%~\cite{Denner} at an electron beam energy of 4.6~GeV. We decided not to correct for this effect and include it in the error budget. Systematic uncertainties on the analyzing power are summarized in Table~\ref{table:systAs}.

\subsection{Photon polarization}
The circular photon polarization is measured at the exit of the Fabry-Perot cavity using an analysis device composed of a quarter-wave plate, a Wollaston prism and two integrating spheres. This device allows a complete polarization measurement through the four Stokes parameters by rotating the quarter-wave plate. In production mode the quarter-wave plate angle is fixed and the spheres only monitor the time variations of the degree of circular polarization (DOCP). The polarization at the center of the cavity where the Compton interaction takes places is deduced from the Stokes parameters, knowing the optical transport matrix of the exit line. This matrix is determined before the installation of the cavity using a dedicated setup where polarization measurements are performed for various orientations of the elliptic polarization of the light. With this method a precision of 0.4~\% is reached including both the modelisation of the transport and the measurement errors. After the cavity is installed, additional effects coming from mirror transmission, birefringence and optical alignment of photon beam must be taken into account. Since in production mode only the DOCP is measured, we use the observed variations and the transport matrix to determine the envelope of possible variations of the polarization inside the cavity. This results in a 0.4~\% systematic error. All
uncertainties are summarized in Table~\ref{table:systPgamma}  The mean value of the DOCP for both laser polarization states is :
\begin{eqnarray}
\label{eq:pgammaLR}
P_{\gamma}^{L}=+99.9\%\pm0.6\% \\
P_{\gamma}^{R}=+99.3\%\pm0.6\% \nonumber
\end{eqnarray}
The photon polarization used for the electron polarization measurement is the average value between the two polarization states :
\begin{equation}
\label{eq:pgamma}
P_{\gamma}=\frac{\omega^{L}P_{\gamma}^{L}-\omega^{R}P_{\gamma}^{R}}{\omega^{L}+\omega^{R}}
\end{equation}
where we took to first order $\omega^{L}=\omega^{R}$.
\section{Results and discussions}
\subsection{General results}
\begin{figure}[t]
\begin{center}
  \scalebox{0.6}{
    \includegraphics{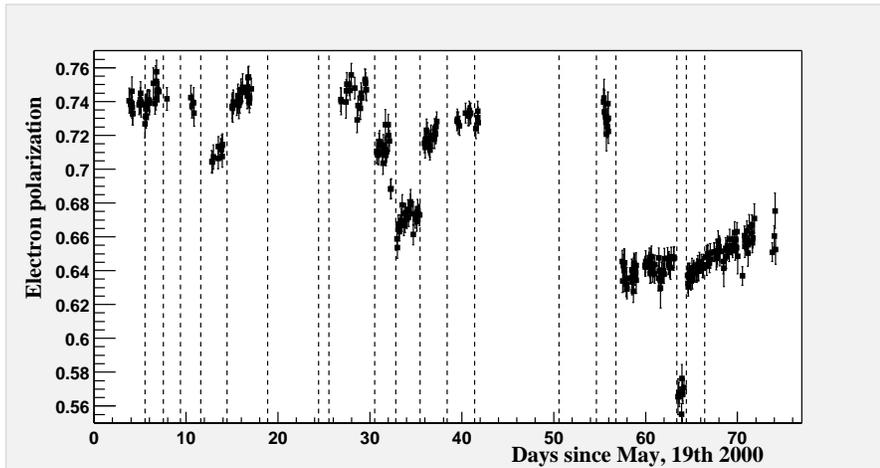}
  }
\end{center}
\caption{\label{fig:ePolar} Electron polarization measurements during N-$\Delta$ experiment. Vertical dash lines show laser spot moves on AsGa crystal at the polarized electron source}
\end{figure}

A review of the uncertainties is given in Table~\ref{table:systT}. The last column shows the accuracy of the monitoring of the electron beam polarization for which all normalization errors cancel. A summary graph of all polarization measurements performed during the N-$\Delta$ experiment is shown in figure~\ref{fig:ePolar} (300 measurements in 60 days). The jumps in the beam polarization are directly correlated with operations at the polarized electron source when the laser spot is displaced to illuminate a different spot on the photocathode in order to increase the beam current. These significant variations in the beam polarization  demonstrate that the Compton polarimeter is an ideal and a mandatory tool to provide a meaningful polarization measurement over long data-taking periods.
\subsection{Determination of $\Delta P_e$}
\begin{figure}[b]
\begin{center}
  \scalebox{0.6}{
    \includegraphics{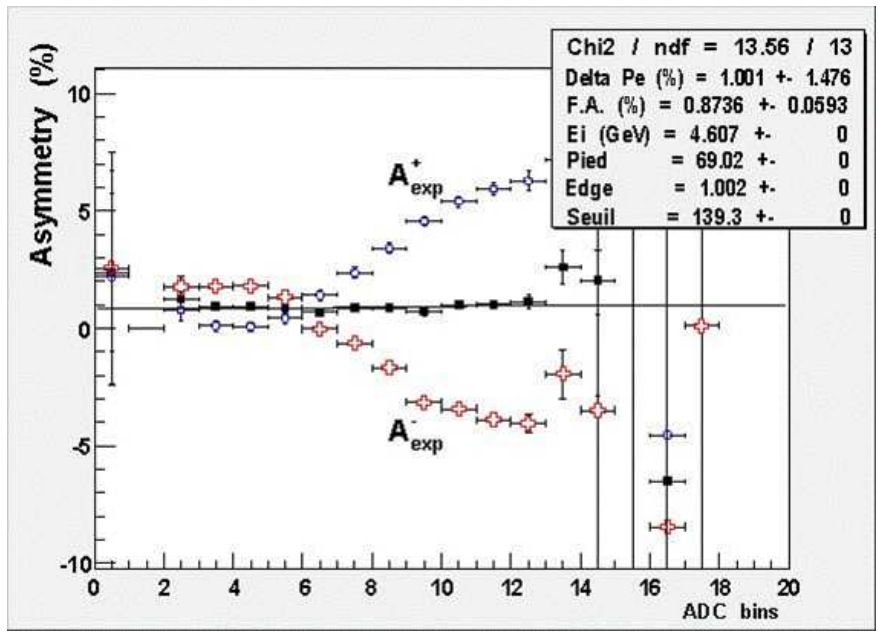}
  }
\end{center}
\caption{\label{fig:DeltaPe} Experimental asymmetries in regard with photon energy, for a positive(+) and negative(-) electron helicity state,
and for the average of both (filled circles).}
\end{figure}

Most of the polarized physics experiments in Hall A are only sensitive to the mean longitudinal electron polarization defined as
\begin{equation}
\label{eq:pe}
P_{e}=\frac{|P_{e}^+|+|P_{e}^-|}{2}
\end{equation}
where $P_{e}^+$ and $P_{e}^-$ denote the electron polarization in each electron spin configuration (parallel or anti-parallel). However, some experiments, such as the  N-$\Delta$ experiment, are sensitive to
\begin{equation}
\label{eq:deltape}
\Delta P_{e}=\frac{|P_{e}^+|-|P_{e}^-|}{2}
\end{equation}
One way to measure this quantity is to use the photon polarization reversal, sacrificing the cancellation of helicity-correlated effects. Experimental asymmetries are thus computed from counting rates between two opposite signs of the photon polarization, for each electron helicity~\cite{Cpt_Steffie}. However, the photon polarization is reversed every three minutes only, resulting in a false asymmetry of the same size as the Compton asymmetry itself. If one makes the assumption that these false asymmetries are independent of the backscattered photon's energy, variation of the Compton asymmetry with respect to energy allows one to isolate $\Delta P_e$. An example is shown in figure~\ref{fig:DeltaPe} where the sum of both experimental asymmetries $A_{exp}^+$ and $A_{exp}^-$ is fitted with a function such as
\begin{equation}
\label{eq:fdeE}
f(E)=\Delta P_{e} \cdot P_\gamma \cdot A_{C}(E)+cst.
\end{equation}
For a set of Left/Right photon reversals over several days, we assess $\Delta P_e$ for the first time at JLab and find it
statistically compatible with zero at a level of 0.3~\%.

\section{Conclusion}
We have continuously measured the CEBAF electron beam polarization over two periods of 30 days at an electron energy of 4.6~GeV and an average current of 40~$\mu$A. The use of a highly segmented electron detector in coincidence with the photon detector was a key element to reduce the systematic errors. By using 40 minute runs a total relative systematic error of 1.2~\% was achieved. Thanks to our high-gain optical cavity and a double beam position feed-back, a statistical accuracy of 1~\% could be reached within 25 minutes. In the relative variations of the beam polarization from one run to another the correlated errors cancel
out and the systematic error is reduced to 0.7~\%. Because most of the recent experiments in Hall A take advantage of the highly polarized and intense electron beam available at JLab, the Compton polarimeter has been routinely operated over the last three years to monitor the beam polarization. Its performance are crucial for the upcoming parity experiments~\cite{helium,hydrogene,plomb} which aim for a very accurate measurements ($\le 2\%$) in an energy range of  0.85 to 3.00~GeV. Such a precision remains challenging and require detectors and laser upgrades which are under study. 
At higher energy (6 GeV), sub-percent measurements are feasible with only minor upgrades of the present apparatus.

\newpage

\begin{table}
\begin{center}
\begin{tabular}{|c|c|}
\hline
type	& error \\ \hline
Background	& 0.05~\% \\
Dead time	& 0.1~\% \\
Beam intensity	& - \\
Events cut	& 0.1~\% \\
position	& 0.3~\% \\ \hline
TOTAL on $<A_{exp}>$	& 0.35~\% \\ \hline
\end{tabular}
\caption{\label{tab:ExpSyst} Run to run systematic uncertainties applied to Compton experimental asymmetry.}
\end{center}
\end{table}

\begin{table}
\begin{center}
\begin{tabular}{|c|c|}
\hline
	& Syst. error \\ \hline
Response function	& 0.45~\% \\
Energy calibration	& 0.6~\% \\
Pile up	                & 0.45~\% \\
Radiative corrections	& 0.26~\% \\ \hline
TOTAL on $<A_s>$	        & 0.95~\% \\ \hline
\end{tabular}
\caption{\label{table:systAs} Relative systematic uncertainties applied to Compton analyzing power during and GEp experiments~\cite{GEp,Gayou}.}
\end{center}
\end{table}

\begin{table}
\begin{center}
\begin{tabular}{|c|c|}
\hline
Time fluctuations	& 0.4~\% \\
Polarization transport	& 0.4~\% \\
Mirrors transmission	& 0.14~\% \\
Birefringence	        & 0.05~\% \\
Alignment	        & 0.1~\% \\ \hline
TOTAL on $P_{\gamma}^{L/R}$  & 0.60~\% \\ \hline
\end{tabular}
\caption{\label{table:systPgamma} Relative systematic uncertainties applied to each  photon beam polarization states.}
\end{center}
\end{table}

\begin{table}
\begin{center}
\begin{tabular}{|c|c|c|}
\hline
                        & Absolute Measurement	 & Monitoring \\ \hline
Experimental asymmetry	& 0.50 \%	& 0.50 \% \\
Analyzing power	        & 0.95 \%	& 0.45 \% \\
Photon polarization	& 0.60 \%	& -       \\
Total systematic	& 1.23 \%	& 0.67 \% \\
Statistical error	& 0.80 \%	&0.80 \%  \\ \hline
TOTAL	                & 1.47 \%	& 1.04 \% \\ \hline
\end{tabular}
\caption{ \label{table:systT} Review of uncertainties for an absolute (2nd column) and relative (3rd column) electron beam polarization measurement.}
\end{center}
\end{table}

\clearpage
\newpage

\newpage

\clearpage
\newpage

\clearpage
\newpage

\clearpage
\newpage

\clearpage
\newpage
\listoffigures

\end{document}